\documentclass{POS}

\usepackage{epsfig,amsmath}

\title{Medium-modified Jets in Heavy-Ion Collisions}

\ShortTitle{Medium-modified Jets}

\author{\speaker{Thorsten Renk}\\
        Department of Physics, University of Jyv\"{a}skyl\"{a}, Finland and Helsinki Institute of Physics, Helsinki, Finland\\
        E-mail: \email{thorsten.i.renk@jyu.fi}}

\abstract{The suppression of single inclusive hadron spectra in heavy-ion collision as compared to the scaled expectation from proton-proton collisions has long been regarded as an interesting tool to study properties of the bulk matter in heavy-ion collisions. However, the limitations of this class of observables has become increasingly obvious, and both experimental and theoretical efforts are now made to go beyond single hadrons to fully reconstructed jets. Monte-Carlo (MC) simulations of in-medium parton showers are currently considered the most promising tool to theoretically access jet physics in heavy-ion collisions. In this paper, I review some of the first results obtained with the MC code YaJEM (Yet another Jet Energy-loss Model) for both single hadron and jet observables.}

\FullConference{High-pT Physics at LHC -09\\
		 February 4- 4 2009\\
		 Prague, Czech Republic}

\begin{document}

\section{Introduction}

Since the suppression of high transverse momentum ($P_T$) hadrons in heavy-ion (A-A) collisions as compared to the scaled expectation from proton-proton (p-p) collisions has been suggested theoretically as a tool to study properties of the hot and dense QCD matter produced in A-A collisions \cite{Jets}, a substantial effort has been made on the experimental side to measure high $P_T$ hadrons. Current A-A data include the suppression of single inclusive hadrons as a function of $P_T$, both averaged \cite{PHENIX-RAA} and as a function of the angle $\phi$ of the hard hadron with the reaction plane and the suppression of  back-to-back angular correlations \cite{STAR-Dihadrons}. 

Single hadron observables and back-to-back correlations are well described in detailed model calculations using the concept of energy loss \cite{HydroJet1,Dihadron1,Dihadron2}, i.e. under the assumption that the process can be described by a medium-induced shift of the leading parton energy by an amount $\Delta E$, followed by a fragmentation process with the vacuum fragmentation function of a parton with the reduced energy. However, there are also calculations for these observables in which the evolution of the whole in-medium parton shower is followed in an analytic way \cite{HydroJet2,HydroJet3,Dihadron3}. A recent comparison study \cite{RAA-Comp} demonstrates both the success of different models in describing the available data on single hadron suppression and the limitation of these observables in terms of constraining properties of the medium, as models with quite different physics assumptions are able to account well for the available data.

One way of resolving this situation is to move from measurements of single hadron spectra to fully reconstructed jets. While the former measurement is sensitive primarily to the leading hadron of the jet from which energy is lost, in the latter measurement also subleading hadrons are observed and hence the redistribution of energy by the medium from the leading hadron to subleading hadrons or into the medium can be traced. On the experimental side, first measurements of jets have become available \cite{STARJET}. MC simulations of in-medium parton showers are considered the appropriate tool to discuss this physics for a number of reasons. One important argument is that explicit conservation of energy and momentum may be a more important factor for jet observables than quantum coherence. Energy momentum conservation is not usually present in analytic calculations as they are commonly done in an asymptotic limit, but is easily accounted for in a MC code. On the other hand, quantum coherence is more easily accounted for in an analytical calculation than in a MC code \cite{JEWEL}. Another crucial factor is that the experimental procedures to reconstruct jets in an heavy-ion environment can be readily applied to the output to a MC simulation, but it is currently unclear how to compare data with analyical calculations in an unbiased way. Thus, at present a number of MC codes to simulate jets in medium are being developed, among them JEWEL \cite{JEWEL}, YaJEM \cite{YaJEM, YaJEM2, YaJEM3} and Q-PYTHIA \cite{Q-PYTHIA}. In this paper, I discuss the physics underlying YaJEM and the modifications to jets in heavy-ion collisions based on the results of the code while making the connection to older results based on the energy loss picture.

\section{Parton shower and fragmentation in vacuum}

In factorized Quantum Chromodynamics (QCD), the production of hadrons from a p-p collision can in leading order be written schematically as

\begin{equation}
d\sigma^{NN \rightarrow h+X} = \sum_{fijk}  f_{i/N}(x_1,Q^2) \otimes  f_{j/N}(x_2, Q^2) \otimes \hat{\sigma}_{ij 
\rightarrow f+k}  \otimes  D_{f\rightarrow h}^{vac}(z, Q_f^2).
\end{equation}

Here, $f_{i(j)/N}(x_{1(2)}, Q^2)$ stand for the initial distributions of parton species $i,j$ with momentum faction $x_{1(2)}$ in the nucleon probed at scale $Q^2$. $\hat{\sigma}_{ij \rightarrow f+k}$ is the perturbatively calculable hard process of a reaction in which partons $i,j$ scatter into partons $f,k$ and $D_{f\rightarrow h}^{vac}(z, Q_f^2)$ is the distribution of hadrons $h$ with momentum fraction $z$  produced from a parton $f$ probed at a scale $\mu_f^2$. Both $f$ and $D$ encode non-perturbative input, in particular the physics of hadronization and confinement, but their scale evolution with $Q^2$ or $Q_f^2$ is perturbatively calculable. Or, in other words, $D_{f\rightarrow h}$ encodes both a perturbative partonic evolution from a parton $f$ at an initial high virtuality scale $Q^2$ into a shower of partons at a lower scale $Q_0^2$ where perturbative QCD ceases to be applicable and a non-perturbative evolution of a parton shower at scale $Q_0^2$ into a hadron shower where the relevant scale is finally set by the squared hadron mass $m_h^2$.

If one wants to study the effect of the hot QCD medium produced in A-A collisions on this expression, it is useful to analyze these virtuality scales. By the uncertainty relation, the relevant timescale at which a fluctuation at a scale $Q$ develops is $1/Q$ in its own restframe, in the lab frame an additional Lorentz factor $E/Q$ appears where $E$ is the energy of the parton (hadron), thus for the evolution time $\tau \sim E/Q^2$. Inserting typical energies $E$ of 10-20 GeV for hard probes at RHIC and an initial virtuality scale $Q_i$ of the same order which is relevant for the hard process itself, one finds timescales $\tau \ll 1$ fm/c, i.e. the hard process occurs long before the thermalization timescale of $\tau \sim 0.6-1$ fm/c at which the medium is produced. On the other hand, inserting the pion mass as a typical hadronic scale into the expression, $Q \sim m_\pi$ , one finds $\tau \gg \tau_{med} \approx 10-20$ fm/c, i.e. the hadronization timescale is much greater than the lifetime of the medium. Thus, it follows that neither the parton distributions (which are an initial state effect relevant before the hard process), nor the hard process itself nor hadronization are affected by the medium, but chiefly the partonic evolution from an initial scale $Q_i$ down to the hadronization scale $Q_0$. It is this evolution a computation of medium effects needs to focus on. 

Parton showers in vacuum can be modelled as a series of branchings $a \rightarrow b,c$ with the three reactions $q \rightarrow qg, g \rightarrow gg$ and $g \rightarrow q\overline{q}$. Conveniently, the evolution is discussed in terms of two evolution variables in momentum space, $t = \ln Q^2/\Lambda_{QCD}$ as a measure of the (decreasing) virtuality scale and $z$ the energy (momentum) splitting variable in each branching as $E_b = z E_a$ and $E_c = (1-z) E_a$. In these variables, the differential branching probability at scale $t$ is then given by

\begin{equation}
\label{E-dP}
dP_a = \sum_{b,c} \frac{\alpha_s(t)}{2\pi} P_{a\rightarrow bc}(z) dt dz
\end{equation}
with the splitting kernels calculable in perturbative QCD as
\begin{equation}
\label{E-Kernel}
P_{q\rightarrow qg}(z) = \frac{4}{3} \frac{1+z^2}{1-z} \quad P_{g\rightarrow gg}(z) = 3 \frac{(1-z(1-z))^2}{z(1-z)} \quad P_{g\rightarrow q\overline{q}}(z) = \frac{N_F}{2} (z^2 + (1-z)^2).
\end{equation}

The probability density that, given an initial scale $t_{in}$, the next branching occurs at a lower scale $t_m$ is then given by

\begin{equation}
\label{E-Qsq}
\frac{dP_a}{dt_m} = \left[\sum_{b,c}I_{a\rightarrow bc}(t_m)  \right] \exp\left[ - \int_{t_{in}}^{t_m} dt' \sum_{b,c} I_{a \rightarrow bc}(t') \right]
\end{equation}

where $I_{a\rightarrow bc}(t_m)$ is Eq.~(\ref{E-dP}) integrated over $z$ subject to kinematical bounds given by the parton virtualities. This expression has two terms, the first one stands for the actual branching probability at $t_m$ whereas the second one, the so-called Sudakov formfactor, stands for the probability that no branching has occurred between $t_{in}$ and $t_m$. Eqs.~(\ref{E-dP},\ref{E-Kernel},\ref{E-Qsq}) are solved in order to find the evolution of a parton shower in vacuum, and they need to be modified in order to compute the effects of the medium on the shower. If the expressions are solved in a MC simulation, quantum coherence can at least be approximately be included by imposing an angular ordering condition

\begin{equation}
\frac{z_b (1-z_b)}{M_b^2} > \frac{1-z_a}{z_a M_a^2}
\end{equation}

at each branching. YaJEM uses the PYSHOW routine \cite{PYSHOW} which is part of the PYTHIA package \cite{PYTHIA} to solve the evolution equations in vacuum and a modifications of this routine to solve them in the medium.

\section{Energy loss as an approximation to in-medium shower evolution}

If one is interested in computing the medium-modification of single inclusive hadron spectra or back-to-back correlations, one can make an approximation in computing the medium-modified fragmentation function. The reason is that the primary parton momentum spectrum is steeply falling with parton momentum $p_T$. Therefore, hadron spectra are dominated by comparatively hard fragmentation, for fragmenting quarks at RHIC kinematics $z \sim 0.7$. This means that one is dominated by events in which a single parton carries most of the momentum of the shower, and the fragmentation function chiefly parametrizes the hadronization of this leading parton and not so much the partonic shower. Any medium effect can then be pictured as energy loss from this leading parton. This is in contrast to an unbiased hard event in which the shower momentum is shared among many partons, thus leading to multiple low $P_T$ hadron production. In the latter situation, clearly the assumption of energy loss from a leading parton is inadequate.

In the energy loss approximation, the key quantity characterizing the medium is the geometry-averaged energy loss probability density $\langle P_f(\Delta E, E) \rangle_{T_{AA}}$ which appears in the factorized computation between hard process and fragmentation. The medium-modified production of hadrons can then be computed (schematically) from

\begin{equation}
d\sigma_{med}^{AA\rightarrow h+X} = \sum_f d\sigma_{vac}^{AA \rightarrow f +X} \otimes{\langle P_f(\Delta E, E) \rangle_{T_{AA}}} \otimes
{D_{f \rightarrow h}^{vac}(z, \mu_F^2)}
\end{equation}
and
\begin{equation}
d\sigma_{vac}^{AA \rightarrow f +X} = \sum_{ijk} {f_{i/A}(x_1,Q^2)} \otimes {f_{j/A}(x_2, Q^2)} \otimes {\hat{\sigma}_{ij 
\rightarrow f+k}}.
\end{equation}
The nuclear suppression factor $R_{AA}$ can in this framework be written as 

\begin{equation}
R_{AA}(P_T) = d\sigma_{med}^{AA\rightarrow h+X}(P_T)/d\sigma^{NN \rightarrow h+X}(P_T).
\end{equation}

\begin{figure}[htb]
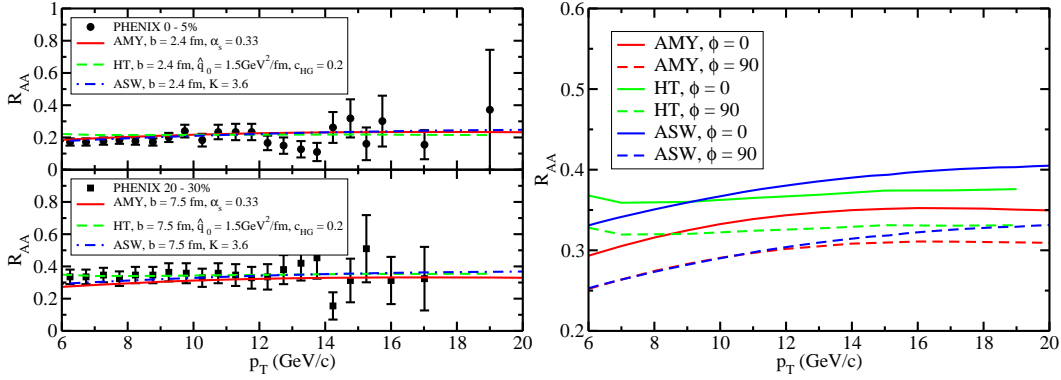

\epsfig{file=RAA_centrality.eps, width=7cm}\epsfig{file=RAA_reaction_plane.eps, width=7cm}
\caption{\label{F-Comp}Left panel: Nuclear suppression factor $R_{AA}$ averaged over the angle with the reaction plane as a function of hadron momentum $P_T$ in 200 AGeV Au-Au collisions in three different framworks compared with PHENIX data for two different centralities. Right panel: $R_{AA}$ calculated as a function of $P_T$ in noncentral 200 AGeV Au-Au collisions assuming an impact parameter $b=7.5$ fm for in plane ($\phi = 0$) and out of plane ($\phi= 90$) emission where $\phi$ is the angle of the outgoing parton with the reaction plane (figures from \cite{RAA-Comp}).}
\end{figure}

The averaging of the energy loss probability density is understood to be computed over all possible paths through the medium with an initial vertex distribution given by the nuclear overlap, i.e. the density of binary collisions. The detailed shape of $P(\Delta E, E)$ given a single path in the medium can be computed in various models. In a recent comparison study \cite{RAA-Comp}, the medium modification of hadron spectra has been computed in three different frameworks with the averaging done over a 3-d hydrodynamical evolution model of the bulk medium. In each model, there is one free parameter which links the strength of the medium modification to the thermodynamical variables in the hydrodynamical model. This parameter is adjusted to data. The result in terms of the nuclear suppression factor $R_{AA}$ is shown in Fig.~\ref{F-Comp}.

It is clear from these results that all models describe both the $P_T$ and the centrality dependence of the data well, although they exhibit some differences in the in plane vs. out of plane suppression which could be used to constrain the physics of the parton-medium interaction more. This similarity of the results, in spite of different physics assumptions entering the models, strongly emphasizes the need to go beyond the energy loss approximation. However, already leading hadron suppression allows some constraints on the physics of parton-medium interaction. In \cite{Elastic}, using the energy loss approximation for both single hadron and back-to-back correlation suppression, it could be shown that the parametric pathlength dependence of the medium modification agrees with the idea that the dominant medium effect is induced gluon radiation, but does not agree with the idea that the parton interacts with the medium chiefly by elastic collisions. Tentatively, the contribution of elastic processes could be constrained to be about 10\%.

\section{In-medium shower evolution and jets}

There are several advantages to studying jets instead of single inclusive hadron spectra, in spite of the experimental difficulty of identifying jets above the event background in a heavy-ion collision. First, jets reflect unbiased hard events in which the momentum flow is not constrained to be through a single parton, hence given the same shower momentum, they can be measured with higher statistics, essentially leading to a larger experimental reach in $P_T$. Furthermore, jet observables allow to observe the energy flow from leading to subleading partons and hence offer a more detailed window on medium effects than energy loss from a leading parton. Finally, shower evolution is usually computed in momentum space, but the medium provides a 'meter stick' in terms of its finite extension and a 'clock' in terms of its finite lifetime to observe the evolution also in position space.

The last property actually makes it necessary for any model of in-medium shower evolution to link the momentum space evolution outlined above with the propagation of partons in position space where the bulk medium is evolved. In the following, I outline the physics assumptions underlying the MC code YaJEM. A detailed description of the model can be found in \cite{YaJEM,YaJEM3}. In YaJEM the link to the spacetime dynamics is made by modelling the average time for a parton $b$ to branch from parent $a$ given the parton energies and virtualities based on the uncertainty relation as

\begin{equation}
\langle \tau_b  \rangle= \frac{E_b}{Q_b^2} - \frac{E_b}{Q_a^2}
\end{equation}  

whereas the actual time in given branching is generated from the exponential branching probability distribution

\begin{equation}
P(\tau_b) = \exp\left[- \frac{\tau_b}{\langle \tau_b \rangle}  \right].
\end{equation}

For simplicity, all partons of a shower are propagated with this time information along an eikonal trajectory determined by the shower initiator. This amounts to neglecting the spread in transverse space when probing the medium.

Currently, YaJEM models three different scenarios for the parton-medium interaction, two of which modify the kinematics of the propagating parton whereas the last modifies the branching probabilities at each vertex (thus, in the first two scenarios the energy of the shower is {\em not} the energy of the shower initiating parton as there is explicit energy transfer between shower and medium, whereas in the last scenario the energy in the shower is conserved). In the RAD scenario, the medium is assumed to cause an increase $\Delta Q_a^2$ in the virtuality of a parton $a$ based on a local transport coefficient $\hat{q}(\zeta)$ as given by the line integral

\begin{equation}
\Delta Q_a^2 = \int_{\tau_a^0}^{\tau_a^0 + \tau_a} d\zeta \hat{q}(\zeta).
\end{equation}

This modification causes medium-induced radiation. Another modification, referred to as DRAG is a drag coefficient $D(\zeta)$ which is motivated by the result of AdS/CFT computations \cite{AdS} of energy loss in a strongly coupled medium. This induces energy loss $\Delta E_a$ for every propagating parton $a$ according to the line integral

\begin{equation}
\Delta E_a = \int_{\tau_a^0}^{\tau_a^0 + \tau_a} d\zeta D \rho(\zeta).
\end{equation}

The last scenario, called FMED, is included for comparison with the modelling done in \cite{JEWEL,HBP}. Here, the singular part of the splitting kernel in the evolution equations is enhanced by a factor $(1+f_{med})$ where the assumption $f_{med} \sim \int d\zeta \rho(\zeta)$ provides the link with the spacetime evolution in YaJEM.

\begin{figure}[htb]
\epsfig{file=D_comp.eps, width=7cm}\raisebox{-2mm}{\epsfig{file=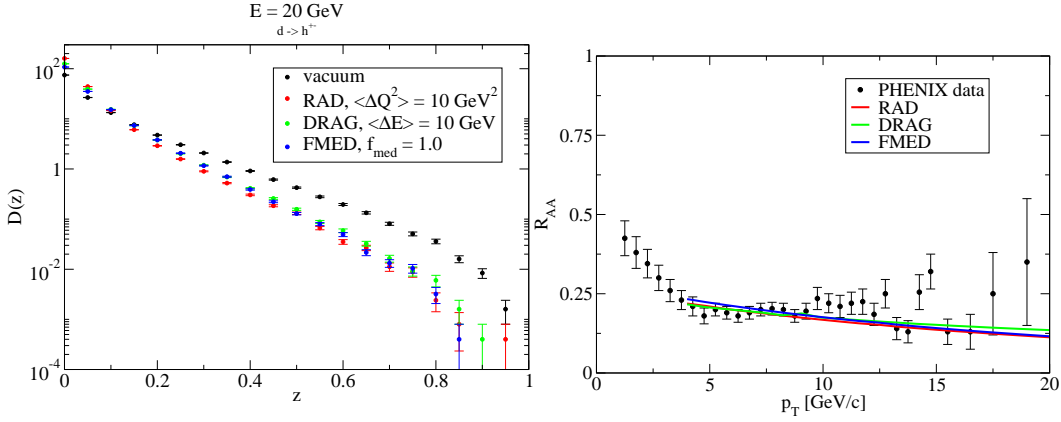, width=7cm}}
\caption{\label{F-YaJEM1}Left panel: Medium-modified fragmentation function of a 20 GeV $d$-quark computed in YaJEM for the vacuum and three different scenarios of parton-medium interaction (see text), adjusted such that the fragmentation functions agree around $z\sim 0.7$. Right panel: Nuclear modification factor $R_{AA}$ for central 200 AGeV Au-Au collisions computed by averaging over a 3-d hydrodynamical model of the medium for three scenarios of parton-medium interaction. }
\end{figure}

Fig.~\ref{F-YaJEM1}, left panel shows the medium-modified fragmentation function for a 20 GeV $d$-quark for a path through the medium in all three scenarios characterized by the specific medium parameters, where these parameters have been selected for the purpose of comparison such that the fragmentation functions agree around $z \sim 0.7$. It is clearly evident that the medium causes a depletion of the fragmentation function at high $z$. This is, in the language of full in-medium shower evolution, the equivalent of energy loss. In the right panel, $R_{AA}$ is computed by averaging over all possible paths through the medium. The medium-induced suppression is clearly visible, although the decreasing trend with $P_T$ is somewhat puzzling. However, the same trend is also observed in other calculations of full in-medium shower evolution.

\begin{figure}[htb]
\epsfig{file=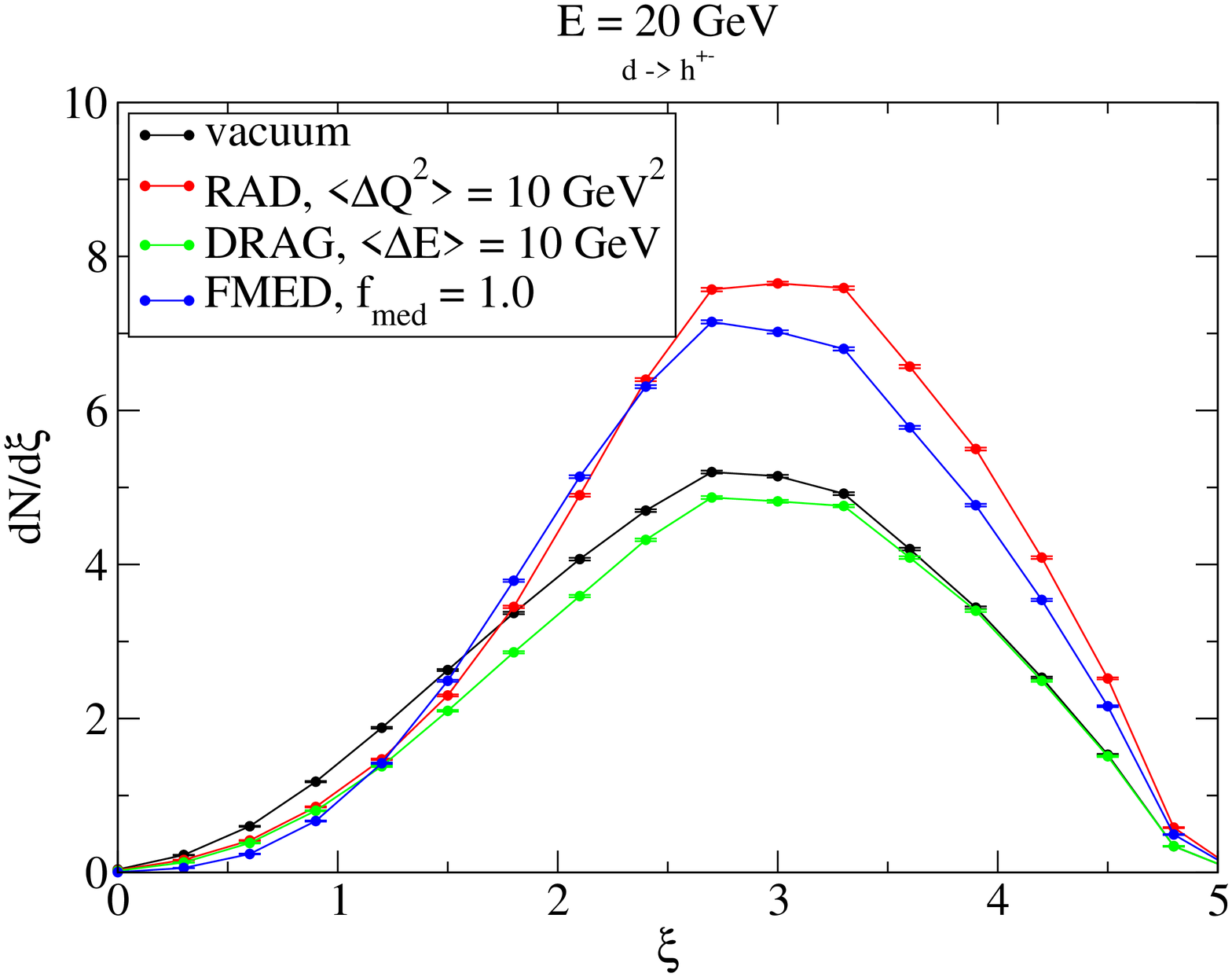, width=7cm}\raisebox{-2mm}{\epsfig{file=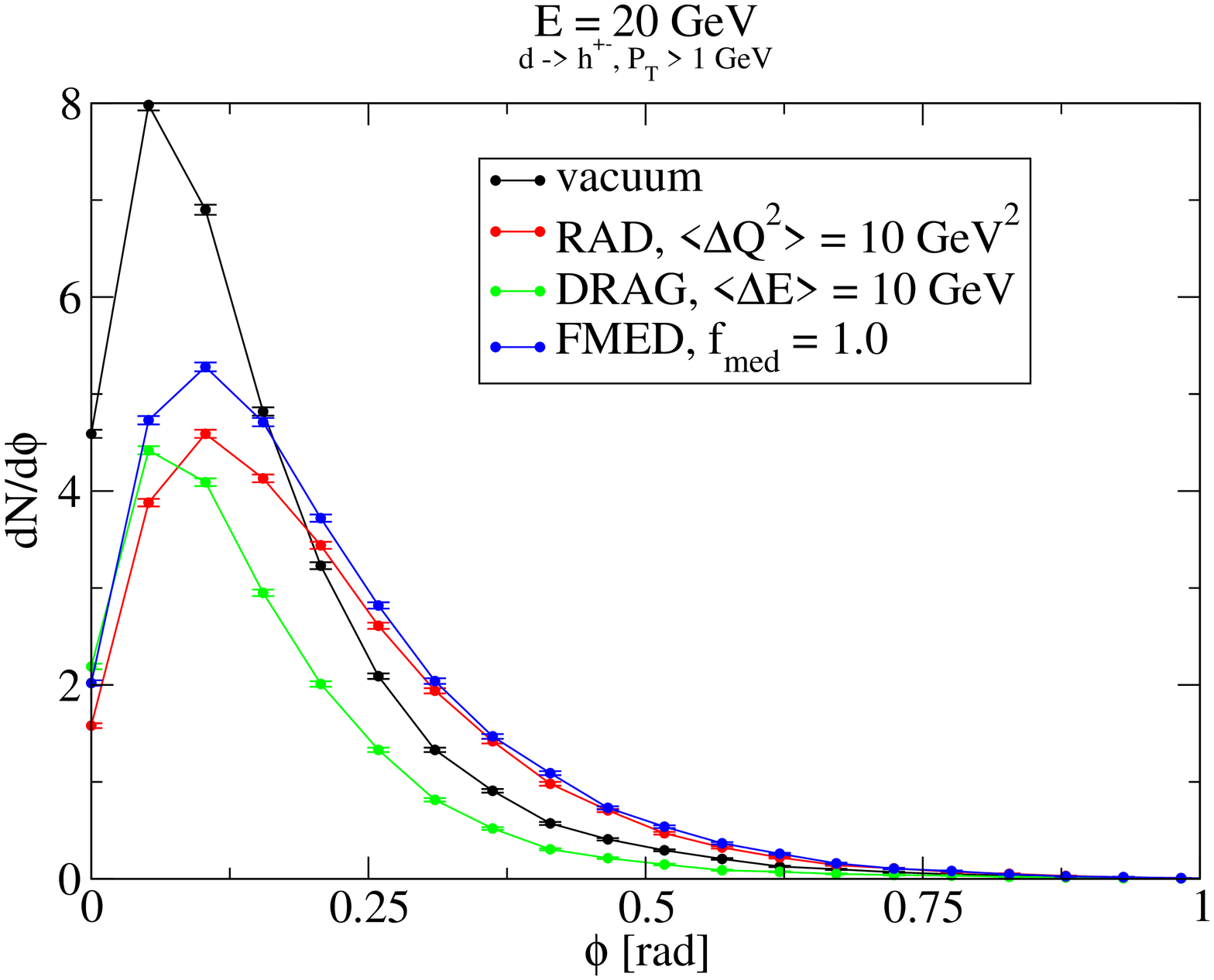, width=7cm}}
\caption{\label{F-YaJEM2}Left panel: $dN/d\xi$ for vacuum and three different scenarios of in-medium shower evolution (see text) Right panel: The angular distribution of hadrons in the shower with respect to the jet axis above 1 GeV momentum the vacuum and three different scenarios of in-medium shower evolution (see text).}
\end{figure}

Based on these results, it would seem difficult to distinguish the different scenarios. However, one can also look at other jet observables, such as the longitudinal momentum distribution inside the jet in terms of $\xi = \ln [1/z]$ which magnifies the low $z$ region of the fragmentation function or the angular distribution of hadrons in the jet. This is shown in Fig.~\ref{F-YaJEM2}.

For the two scenarios generating medium-induced radiation, RAD and FMED, a clear enhancement in the high $\xi$ region over the vacuum result is visible in $dN/d\xi$, whereas this is not seen for the DRAG result. This corresponds to the presence or absence of multiple low $P_T$ hadron production in the hadronization of the medium induced radiation. A measurement of this observable could distinguish between a radiative energy loss mechanism and a drag force which are able to produce a near-identical $R_{AA}$.

A similar case can be made with the angular distribution of hadrons, depicted in Fig.~\ref{F-YaJEM2}, right panel. Here, the radiative scenarios RAD and FMED lead to a widening of the jet whereas the DRAG scenario does not. If this could be measured, it could be a handle on the physics process underlying energy the parton-medium interaction.

There is, however, an important caveat: Any experimental measurement of jets in a heavy-ion environment will have to make use of cuts to identify the jet. These cuts in turn will cause a systematic bias on the jet sample that is studied. In \cite{YaJEM2} it was shown that if the set of cuts is chosen too tightly, the resulting bias makes the medium modification of an observable (in the paper $dN/d\xi$ was considered)  insignificant.

\section{Summary}

In this paper, I have presented arguments that jet observables contain information beyond what can be gained from a study of single inclusive hadron spectra and dihadron correlations in heavy-ion collisions. From arguments based on the uncertainty relation, one can deduce that the medium modification of a hard process chiefly affects the partonic evolution of a shower. For observables sensitive to the leading hadron of a jet only, the dominant kinematics is such that a single parton carries most of the momentum in the shower. As a consequence, the in-medium evolution of the shower can be modelled as energy loss from the leading parton. While computations based on this approximation are very successful when confronted with data, it remains difficult to distinguish different physics assumptions underlying models. An exception to this is the pathlength dependence of energy loss which can be constrained well from the suppression of back-to-back correlations and which agrees with medium-induced radiation as the main mechanism of energy loss, but not with elastic collisions.

Jet observables have the potential to reveal more about the nature of parton-medium interaction, as they trace the redistribution of energy in the medium. Since explicit energy-momentum conservation may be the most important factor in understanding this redistribution, the theoretical investigation of these observables is best made by MC simulations of the in-medium parton shower evolution. In this paper, I have outlined the physics underlying the MC shower code YaJEM. First results applying the code to single hadron observables or jet propertis find indeed the expected result --- while the nuclear suppression factor $R_{AA}$ is unable to distinguish between medium-induced raditation and a drag force as the effect causing the medium modification, jet observables such as the angular distribution of hadrons with respect to the jet axis or the longitudinal distribution of momenta in the jet show clear differences. 

There is now good reason to assume that jet observables will reveal important information about the microscopical properties of the medium, and Monte Carlo simulations of in-medium showers such as YaJEM will most likely be the appropriate tools to extract this information.

\end{document}